\documentclass[prl,twocolumn,superscriptaddress,floatfix,10pt]{revtex4-1}
\usepackage{graphicx,bm}
\usepackage{amsmath}
\usepackage{amsfonts}
\usepackage{amssymb}
\usepackage{braket}
\usepackage{dsfont}
\usepackage{bm}
\usepackage{eucal}

\usepackage{mathtools}

\usepackage{color}

\usepackage{xcolor}
\usepackage[colorlinks=true,urlcolor=blue,linkcolor=blue,citecolor=blue]{hyperref}
\usepackage{cleveref} %should be placed only after hyperred

\usepackage{comment}
\usepackage[normalem]{ulem}

\usepackage{tikz}
\usetikzlibrary{calc}
\usetikzlibrary{decorations.markings}
\usetikzlibrary{decorations.pathmorphing, arrows}
\usetikzlibrary{arrows.meta}

\usepackage[makeroom]{cancel}

\usepackage{pdfpages}
\usepackage{pgffor}
\makeatletter
\AtBeginDocument{\let\LS@rot\@undefined}
\makeatother

\begin{document}
\title{Superdiffusive transport in chaotic quantum systems with nodal interactions}

\author{Yu-Peng Wang}
\thanks{These authors contributed to this work equally.}
\affiliation{Beijing National Laboratory for Condensed Matter Physics and Institute of Physics, Chinese Academy of Sciences, Beijing 100190, China}
\affiliation{Institute of Science and Technology Austria (ISTA), Am Campus 1, 3400 Klosterneuburg, Austria}
\author{Jie Ren}
\thanks{These authors contributed to this work equally.}
\affiliation{Beijing National Laboratory for Condensed Matter Physics and Institute of Physics, Chinese Academy of Sciences, Beijing 100190, China}
\affiliation{School of Physics and Astronomy, University of Leeds, Leeds LS2 9JT, United Kingdom}

\author{Sarang Gopalakrishnan}
\affiliation{Department of Electrical and Computer Engineering,Princeton University, Princeton, NJ 08544, USA}

\author{Romain Vasseur}
\affiliation{Department of Theoretical Physics, University of Geneva, 24 quai Ernest-Ansermet, 1211 Gen\`eve, Switzerland}

\begin{abstract}

We introduce a class of interacting fermionic quantum models in $d$ dimensions with nodal interactions that exhibit superdiffusive transport. We establish non-perturbatively that the nodal structure of the interactions gives rise to long-lived quasiparticle excitations that result in a diverging diffusion constant, even though the system is fully chaotic. Using a Boltzmann equation approach, we find that the charge mode acquires an anomalous dispersion relation at long wavelength $\omega(q) \sim q^{z} $ with dynamical exponent $z={\rm min}[(2n+d)/2n,2]$, where $n$ is the order of the nodal point in momentum space. We verify our predictions in one dimensional systems using tensor-network techniques. 

%In this study, we construct chaotic superdiffusive models using free fermions and nodal interactions. 
%The nodal interaction approximately preserves $\hat n_{k}$ near specific momenta, referred to as nodes. 
%This nodal structure leads to a asymptotic divergence in particle lifetime near nodes, which then results in superdiffusion.
%By analyzing the quantum Boltzmann equation, we find that the near-equilibrium dynamics of this model can be described as a L\'evy walk process, and the dynamical exponent is $z=(2n+1)/2n$, where $n$ is the order of the node.
\end{abstract}

\maketitle

\textit{Introduction.---}
The study of the emergence of hydrodynamic behavior and of the transport of conserved quantities such as charge, spin or energy in many-body quantum systems has attracted significant interest in recent years~\cite{EFT, GHD0,GHD00,Lucas_2018,annurevScaffidi, longrange, Zeiher22, Wienand2024,doyon2023generalizedhydrodynamicsperspective}. Generic interacting lattice systems with short range interactions exhibit diffusive transport at finite temperature, as expected from general hydrodynamic principles, see {\it e.g.}~\cite{Marko12,Meisner14,Rosch14,Varma16,Sachdev17,Enej20,RMP21,Romain20,Zeiher22}, and finding generic deviations from diffusion is an important challenge. Slower than diffusive (subdiffusive) transport can naturally occur as a result of disorder and localization phenomena~\cite{anderson1958,Varma16,Ljubotina18,Reichman15,Demler15,Altshuler06,Karrasch17}, or because of kinetic constraints~\cite{sub2,sub6,sub4,sub5,sub1,PhysRevLett.127.230602,PhysRevResearch.4.L012003,sub3,Fredkin}. On the other end of the spectrum, integrable systems support stable quasi-particle excitations and generically show ballistic transport. 

Superdiffusive transport -- between diffusive and ballistic -- is however particularly elusive.
%All  superdiffusive models studied in previous works belong to three categories: (1) systems with long-range interaction, (2) total momentum-conserving systems, or (3) integrable/non-interacting systems.
In systems with long-range interactions, superdiffusion can naturally occur in the form of L\'evy flights~\cite{Knap20,longrange,longrange1,longrange2,longrange3,longrange4,longrange5}. Momentum-conserving systems can also exhibit superdiffusion in one dimension~\cite{Prosen2000,Narayan2002,Spohn,Prosen2003,Prosen2005}. However, in short-range lattice models, superdiffusion seems to require integrability:
In non-interacting systems, superdiffusion has been shown to emerge for specific types of disorder~\cite{Fibonacci1,Fibonacci2,Fibonacci3,randdimmer,Cane2021,Dhar2001,Harmonic1971,Titov16, Wang24, Junaid24} or dephasing~\cite{Wang23, Marko24}. In interacting systems, superdiffusion, along with Kardar-Parisi-Zhang (KPZ)-like scaling functions, has been observed in integrable models with non-Abelian symmetries~\cite{znidaric11,prosen17,GHD1,prosen19,GHD2,GHD3,annurev24,RMP21}. This superdiffusive behavior appears to be remarkably stable to symmetry-preserving integrability breaking perturbations~\cite{GHD4,PhysRevB.110.L180301,PhysRevLett.133.256301}, but the ultimate fate of transport is believed to be diffusive away from integrable points. 
To our knowledge, the only potential example of a noningrable lattice model exhibiting superdiffusion came from a recent numerical study~\cite{Papic23} that indicates superdiffusive energy transport in PXP and related models up to long time scales: a theoretical explanation for this phenomenon is still lacking.
%Superdiffusion in quantum chaotic (non-integrable) lattice remains largely unexplored, and it is widely believed that quantum chaotic lattice are always diffusive~\cite{RMP21}. However, recent numerical simulations have uncovered superdiffusive energy transport in the PXP model~\cite{Papic23}, although a theoretical explanation for this phenomenon is still lacking.

%------------

In this work, we introduce a systematic method for constructing chaotic (non-integrable) lattice models with superdiffusive charge transport. Motivated by the anomalous properties of non-interacting systems with a nodal structure~\cite{Wang23,Wang24}, we construct {\em interacting}, chaotic superdiffusive models from two key elements. The first ingredient is a free fermion Hamiltonian $\hat H_0 = \sum_k \epsilon(k) \hat n_k$, that supports stable quasiparticle excitations and ballistic transport. The second ingredient is a nodal interaction $\hat V = \sum_i \hat V_i$, where all local interaction terms $\hat V_i$ satisfy $[\hat V_i, \hat n_{k_0}] =0$ for a specific momentum $k_0$. The interaction couples quasi-particles with different momenta, and generically leads to thermalization and chaotic behavior. However, due to the nodal structure $[\hat V_i, \hat n_{k_0}] =0$, the quasiparticle lifetime diverges asymptotically  near momentum $k_0$, leading to anomalous transport. For such interacting systems, $\hat n_k$ is approximately conserved as $k \to k_0$ up to very long times. Using a time-dependent Mazur-like bound~\cite{MAZUR1969533}, we obtain a divergent lower bound for the diffusion constant, thereby establishing superdiffusion non-perturbatively. We also characterize transport using a Boltzmann equation approach, and find that the charge mode is governed by a universal dynamical exponent $z = (2n+1)/2n$ where $n$ is the order of the node. We verify our results numerically using tensor-network techniques.

\textit{Models with nodal interactions.---} We consider fermionic systems in $d$ dimensions subject to interactions with a nodal structure. More precisely, we construct local operators $\hat V_i$ that do not affect quasi-particles with a specific momentum $k_0$, i.e. $[\hat V_i, \hat n_{k_0}]=0$, with $\hat n_{k}=\hat c^\dagger_{k} \hat c_{k}$. An important observation is that $[\sum_a \phi_a \hat c_{m+a}, \hat n_k] = e^{ikm}\hat c_k\sum_a \phi_a e^{ika}$, which implies that the operator $\hat d_m \equiv \sum_a \phi_a \hat c_{m+a}$ commutes with  $\hat n_{k_0}$ for all site $m$, provided the Fourier transform of its coefficient, $\tilde{\phi}_k \equiv \sum_a \phi_a e^{ika}$, has a node at $k_0$, i.e. $\tilde{\phi}_{k_0}=0$. Building on this observation, we can construct interaction terms $\hat{V}_i$ from the operators $\hat{d}_i$s and $\hat{d}_i^\dagger$s, so modes with momenta $k_0$ are unaffected by this interaction. As we will demonstrate in this letter, this structure is sufficient to enforce superdiffusive transport. 

\begin{figure*}
	\centering
	\includegraphics[width=\linewidth]{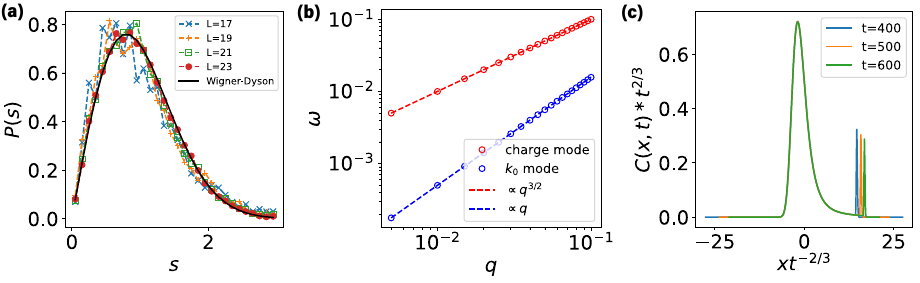}
	\caption{{\bf Superdiffusion in chaotic nodal chains}. (a) Distribution of many-body level spacing $s$ in the middle half of the spectrum of nodal interaction model with $\tilde{\phi}_k=1+e^{i(k+\pi/2)}$. In this figure, we choose momentum $k=0$, particle number $N=(L+1)/2$ and interaction strength $W=4$.  The $r$ ratio~\cite{PhysRevB.75.155111} is consistent with Wigner-Dyson GOE. There is clear level repulsion in this model, ruling out integrability. (b) Low-energy spectrum (real part) of normal modes of the linearized Boltzmann equation for nodal interactions with $\tilde{\phi}_k = 1+e^{i(k+\pi/2)}$, exhibiting a node of order $n=1$ at $k_0 = \pi/2$. We choose the interactions to be noisy in time ($\gamma(k) = \beta(k,k') = 1$) to break energy conservation and focus on charge transport. We find two gapless modes: a ballistic mode corresponding to $n_{k_0}$ (which is exactly conserved by the interactions), and a hydrodynamic charge mode with dispersion relation $\omega \sim q^{3/2}$ with $q$ the momentum, indicating superdiffusive transport with dynamical exponent $z=3/2$. (c) Structure factor $C(x,t)\equiv\langle n(x,t)n(0,0)\rangle$ in same setup as (b), obtained from the Boltzmann equation. The main component of the structure factor follows the scaling relation $C(x,t)=t^{-2/3}f(xt^{-2/3})$, consistent with the dynamical exponent $z=3/2$. The decreasing ballistic component originates from modes near the node $k_0$.}
	\label{fig:fig1}
\end{figure*}

%In non-interacting systems, a previous study\cite{Wang23} employed $\hat L_i= \hat d_i^\dagger \hat d_i$ as the Lindblad operator to obtain superdiffusive dephasing models. Another study\cite{Wang24} used $\hat V_i=\epsilon_i \hat d_i^\dagger \hat d_i$, where $\epsilon_i$ is a random number, to construct superdiffusive disordered models. 
A similar nodal structure was employed to construct non-interacting noisy~\cite{Wang23,Marko24} and disordered systems~\cite{Wang24,Junaid24} with anomalous properties. In this letter, we generalize this construction to {\em interacting} systems, enabling the construction of chaotic (non-integrable) superdiffusive systems. Our approach is very general and can be used in any dimension, but for the sake of simplicity we will  focus on the following nodal one-dimensional spinless fermionic system
\begin{align}\label{eq:Ham}
H = \sum_{k} \epsilon(k) \hat c^\dagger_{k} \hat c_{k} + \sum_{i} W_i(\hat d^\dagger_i \hat d_i \hat d^\dagger_{i+1} \hat d_{i+1}+{\rm h.c.}).
\end{align}
We will also set $W_i =W$, although we note that since the interaction strength $W_i$ does not influence the nodal structure $[\hat V_i, \hat n_{k_0}] =0$, our conclusions will also generalize to inhomogeneous interaction strength $W_i$. This model is generically non-integrable (chaotic), and its level statistics follows a Wigner-Dyson distribution (Fig.~\ref{fig:fig1}a).

Before establishing superdiffusion in this model, we provide an intuitive understanding of its origin. 
Consider a system initially prepared as a slater determinant of $N$ quasi-particles with distinct momenta: $|k_1, \cdots, k_N\rangle$. When interactions are turned on, those states are obviously not eigenstates of the Hamiltonian anymore, and thermalize. However, since $[\hat n_{k_0}, \hat V_i]=0$, the lifetime of quasi-particles with momenta near $k_0$ diverges asymptotically. In the long-time limit, transport is dominated by the surviving  quasi-particles, which carry charge ballistically. This hierarchy of long-lived quasi-particle excitations is responsible for  superdiffusive behavior.

\textit{Divergent diffusion constant.---}
To make this argument more precise, we first establish the divergence of the diffusion constant, corresponding to superdiffusive transport. The diffusion constant at infinite temperature can be calculated from the current-current correlation function from the Kubo formula as
$D = 4\lim_{t\to\infty}\lim_{L\to\infty} \frac{1}{L}\int_{0}^{t}\mathrm{d}\tau \langle \hat J(\tau)\hat J(0)\rangle$~\cite{Sutherland1990,Zhang1992,RMP21},
where $\langle\cdots\rangle = 2^{-L}\mathrm{Tr}(\cdots)$, and $\hat J(t)$ is a sum of local currents over all lattice sites $\hat J(t) = \sum_r \hat j_r(t)$.
We aim to show that the diffusion constant diverges in fermionic systems with nodal interactions.

The main idea is that since $n_k$ is conserved at the node $k_0$, we can treat $n_{k}$ as approximately conserved for momenta in the neighborhood $\mathcal{K}=(k_0-\delta k, k_0+\delta k)$ where $\delta k$ will be time-dependent. To determine the size of the subset $\mathcal{K}$, we use the following bound~\cite{PhysRevE.92.012128} on the infinite-temperature autocorrelation of an arbitrary operator: 
\begin{equation}\label{banuls}
\left \langle (\hat O(t) - \hat O(0)) \hat O(0) \right\rangle \leq \left\langle [\hat H, \hat O][\hat H, \hat O]^\dagger \right\rangle t^2/2.
\end{equation}
We apply this bound~\eqref{banuls} to the operator $\hat n_k$.
%where $\Vert \hat A \Vert_2 \equiv \sqrt{\mathrm{Tr}(A^\dagger A)}$ is the Frobenius norm of the operator.
We show by explicit calculation that 
%
%first note that the evolution of $n_{k}$ is governed by 
%$\partial_t \hat n_{k} = i[\hat H,\hat n_{k}]$, where the operator norm 
%$2^{-L} 
$\left\langle [\hat H, \hat n_k] [\hat H, \hat n_k]^\dagger \right\rangle \leq c^2 |\tilde{\phi}_k|^2$ for some $O(1)$ constant $c$~\cite{suppmat}.
%\le c|d_{k}|$,
%where $c$ is a constant  
%This follows because the terms in $\hat H$ that do not commute with $\hat n_k$ are all proportional to $|d_k|$ and are bounded in their operator norm.  
We consider $\hat n_{k}$ to be approximately conserved during this time interval $(0,t)$ if the right-hand side of Eq.~\eqref{banuls} is small throughout the interval. 
%$|\hat n_{k}(\tau) - \hat n_{k}(0)| < \epsilon$ holds for all $\tau \in (0,t)$, where $\epsilon$ is an arbitrary small number. 
Given a node of order $n$, {\it i.e}. $\tilde{\phi}_k\sim |k-k_0|^n$, for arbitrarily large $t$, as long as momenta are chosen such that $|k-k_0|^nt<O(\sqrt{\epsilon})$, $n_k$ is approximately conserved up to accuracy $\epsilon$. This leads to the scaling $\delta k(t) \sim t^{-1/n}$: for an arbitrary long time $t$, occupation numbers with momenta within $\delta k(t)$ of a node are approximately conserved. % (up to accuracy $\epsilon$).

We can then decompose the current into a slow component, $\hat J_{s}$,
which remains conserved up to time $t$, and a fast component, $\hat J_{f}(t) = \hat J(t) - \hat J_{s}$ which decays quickly and contributes to a finite conductivity. The slow part is given by the (hydrodynamic) projection of the current onto the occupation numbers $\hat n_{k}$ with $k \in \mathcal{K}(t)=(k_0-\delta k (t) , k_0+\delta k (t))$. We have $\hat J_{s} = \sum_{k \in \mathcal{K}(t)}  g(k) \hat n_k $ where the coefficients $g(k)$ are the overlaps of the current onto the occupation numbers -- for weak interactions, we have $g(k) = v(k) + O(W)$ with $v(k) = \partial_k \epsilon(k)$. By definition, the slow part of the current is conserved up to times $t$: $J_s(\tau) \approx J_s(0) $ for $ \tau \leq t$ (up to accuracy $\epsilon$). 
Since the fast and slow components are orthogonal, we can derive a Mazur-like lower bound for the diffusion constant~\cite{suppmat}:
\begin{align}\label{eq:lower bound}
    D \ge 4\lim_{t\to\infty}\lim_{L\to\infty} \frac{1}{L}\int_{0}^{t}\mathrm{d}\tau \langle \hat J_{s}^2\rangle \equiv D_s.
\end{align}
 Our goal is to then show that this lower bound diverges. Using the explicit form of the slow part of the current, we have $ \lim_{L\to\infty} \frac{1}{L} \langle \hat J_{s}^2\rangle = \int_{k \in \mathcal{K}(t)} \frac{dk}{ 2\pi} g(k)^2 n_k(1-n_k) $, which plays the role of a time-dependent Drude weight (note that this quantity does not depend on $\tau$). For sufficiently long time $t$, $\delta k (t)$  becomes very small, so we have $\lim_{L\to\infty} \frac{1}{L} \langle \hat J_{s}^2\rangle \underset{t\to \infty}{\sim} \delta k(t) \frac{1}{ 2\pi} g(k_0)^2 n_{k_0} (1-n_{k_0})$. This gives
\begin{align}
\begin{aligned}
	D_s \sim  g(k_0)^2 \lim_{t\to\infty} t \delta k(t) ,
\end{aligned}
\end{align}
with $\delta k(t) \sim t^{-1/n} $.
This shows that the lower bound $D_s$ diverges for any $n>1$ as long as long as the overlap of the current with the node occupation does not vanish: $g(k_0) = v(k_0) \neq 0$ (see Boltzmann approach below, this is satisfied unless the velocity $v(k_0)$ vanishes at the node).
Therefore, the diffusion constant diverges in the nodal interaction model, and transport must be superdiffusive, at least for $n>1$. %A more careful analysis can yield a tighter lower bound, extending this divergence to $n=1$\cite{suppmat}. %indicating the presence of superdiffusive transport.

\textit{Boltzmann Equation.---}
Although the argument above establishes superdiffusion in our model in a non-perturbative way, it does not predict the dynamical exponent $z$ associated with charge transport. More precisely, the argument above can be turned into a lower bound for an effective time-dependent diffusion constant $t^{1-1/n}  \lesssim D(t)$, but the actual diffusion constant diverges faster with time (corresponding to a smaller exponent $z$): the $O(\tilde{\phi}_k)$ corrections to the commutator add up incoherently, not constructively.
To estimate the true transport exponents we turn to a Boltzmann equation approach.
%
%In order to characterize transport, we now turn to a Boltzmann equation approach. 
First, we write the Hamiltonian (Eq.~\ref{eq:Ham}) in momentum space
$H = \int \frac{\mathrm{d}k}{2\pi} \epsilon(k) \hat c^\dagger_k \hat c_k  + \int\frac{\mathrm{d}k^4}{(2\pi)^3} \delta_{2\pi}(\underline k)U_{k_1,k_2,k_3,k_4}  \hat c_{k_1}^\dagger \hat c_{k_3} \hat c_{k_2}^\dagger \hat c_{k_4} $ where $U_{k_1,k_2,k_3,k_4} = W \tilde{\phi}_{k_1}^* \tilde{\phi}_{k_2}^* \tilde{\phi}_{k_3} \tilde{\phi}_{k_4} \cos(k_2-k_4)$ and $\delta_{2\pi}(\underline k)=\delta(k_1+k_2-k_3-k_4 \mod 2\pi) $ . The corresponding Boltzmann equation has the general form
$
\partial_t n(x, k, t) + v(k)\partial_x n(x, k, t) = f^{\rm col}_{k}[n]
$,
where $n(x,k,t)$ represents the coarse-grained particle density at position $x$ with momentum $k$, $v(k) = \partial_k \epsilon(k)$ and $f^{\rm col}_{k}(n(x,k,t))$ denotes the collision integral term due to the interactions.

Crucially, the nodal structure $\partial_t \hat n_{k_0} = 0$ implies that the collision integral $f^{\rm col}_{k_0} = 0$. In turn, the nodal form of the interactions gives rise to diverging lifetimes for modes with momenta near $k_0$. 
Linearizing the Boltzmann equation around a thermal equilibrium state, and expanding the distribution function as $n(x,k,t) = n_{\rm eq} + \delta n(x,k,t)$, the general form of the Boltzmann equation for $m$-particle scattering is
%To see this, we linearize the Boltzmann equation. In thermal equilibrium at infinite temperature,  $f^{\rm col}[ n_{\rm eq}] = 0$ and $n_{\rm eq}$ is independent of both $x$ and $k$. For small deviations from equilibrium, $n(x,k,t) = n_{\rm eq} + \delta n(x,k,t)$, the linearized collision integral for our model evaluated perturbatively from Fermi's golden rule can be written in the following general form:
\begin{align}
\begin{aligned}\label{eq:linearized QBE}
    f^{\rm col}_{k}[\delta n] &=  -\gamma(k)|\tilde{\phi}_k|^2 \delta n(x,k,t)\\
    +&\gamma(k)|\tilde{\phi}_k|^2 \int  \mathrm{d}k'\ \beta( k,k')\delta n(x, k', t),
\end{aligned}
\end{align}
where $\gamma(k)$, $\beta(k,k')$ are non-negative, and $\int \mathrm{d}k' \beta( k,k')=1$. The specific form of $\gamma(k)$ and $\beta( k,k')$ are unimportant. The crucial point is that the matrix element involving momentum $k$ for the scattering processes in the nodal problem all go as $\sim |\tilde{\phi}_k| $. Therefore, the collision integral (decay rate) goes as $\sim |\tilde{\phi}_k|^2 $. Examples of specific collision integrals are discussed in the supplemental material~\cite{suppmat}. %For example, 
%(Umklapp) two-body scattering processes give a contribution of the form
% $   f^{\rm col}[n] = \frac{4\pi}{\hbar} \int\frac{\mathrm{d}k^4}{(2\pi)^3}\delta_{2\pi}(\underline k)\delta(\underline\epsilon)|\tilde %U_{k_1,k_2,k_3,k_4}|^2 
%     \delta(k-k_1)[(1-n_{k_1})(1-n_{k_2})n_{k_3}n_{k_4}
%    -(1-n_{k_3})(1-n_{k_4}) n_{k_1}n_{k_2}]$,
%where $n_k$ is a shorthand of $n(x, k, t)$, $\tilde %U_{k_1,k_2,k_3,k_4}= U_{k_1,k_2,k_3,k_4}-U_{k_1,k_2,k_4,k_3}$ and $\delta(\underline\epsilon) = \delta(\epsilon_{k_1}+\epsilon_{k_2}-\epsilon_{k_3}-\epsilon_{k_4})$. More general $m$-body scattering terms can also be expressed in a similar way as Eq.~\eqref{eq:linearized QBE} (see supplementary material for details), and all satisfy the general form~\eqref{eq:linearized QBE}.
\begin{figure*}
	\centering
	\includegraphics[width=\linewidth]{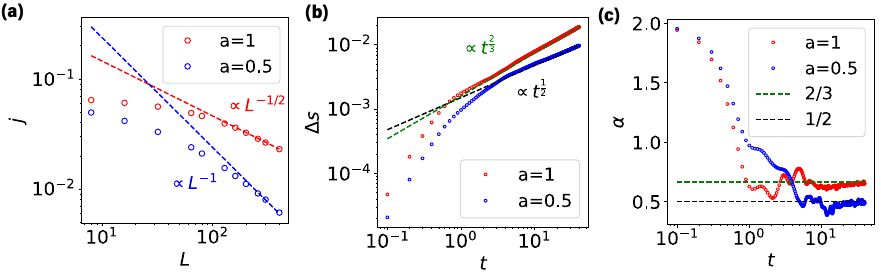}
	\caption{{\bf Tensor-network simulations.}(a) NESS current under boundary driving. When the Hamiltonian possesses node ($a=1$), the scaling of NESS current and system size satisfies $j\sim L^{-1/2}$ at large $L$ limit. In the case $a=0.5$, the interactions do not exhibit a node, and the scaling of NESS current and system size satisfies $j\sim L^{-1}$ at large $L$ limit consistent with diffusive transport. (b) Time dependence of the charge transfer $\Delta s$ starting from a domain wall initial state. (c) The time-dependent exponent $\alpha(t)$ calculated as the numerical logarithmic derivative $\mathrm{d} \log \Delta s/\mathrm{d} \log t$. Fig.~(b) and (c) demonstrate that in the model with nodal interaction, $\Delta s \propto t^{2/3}$, corresponding to a dynamical exponent $z = 3/2$. In contrast, with non-nodal interaction, $\Delta s \propto t^{1/2}$, indicating a dynamical exponent $z = 2$. For the NESS current simulations (a), we chose $\Gamma=1$, $\mu=0.02$, and a maximum bond dimension $\chi = 180$. In the charge transfer simulations (b) and (c), we chose $\nu = 0.01$ and a maximum bond dimension $\chi = 200$. }
	\label{fig:fig2}
\end{figure*}

{\it Long-lived normal modes.---} The general structure of the linearized Boltzmann equation is similar to the case of free fermions with dephasing noise with a nodal structure, studied in Refs.~\cite{Cao19,Wang23}. The linearized Boltzmann equation has a natural interpretation as a Markov process in momentum space: when $\tilde{\phi}_k \ne 0$ for all momenta $k$, the mean free path $l_k = v(k)/(\gamma(k) |\tilde{\phi}_k|^2)$ is finite for all momentum $k$. In this case, the charge transport is diffusive. However, when $\tilde{\phi}_{k_0+q} \sim q^n$ for certain nodes $k_0$, the mean free path diverges near these nodes, resulting in a fat-tailed distribution for the mean-free path $p(\ell) \sim \ell^{-2-1/(2n)}$. This is the hallmark of a L\'evy walk, so charge transport becomes superdiffusive, with the corresponding dynamical exponent given by $z=(1+2n)/2n$. This anomalous scaling can be observed directly in the spectrum of normal modes of the linearized Boltzmann equation. In addition to the exactly ballistic conserved modes $n_{k_0}$, we find a normal mode corresponding to $\sum_{k \neq k_0} n_k$ with dispersion relation
\begin{align}
\omega(q) \sim q^{(1+2n)/2n},
\end{align}
see Fig.~\ref{fig:fig1}(b), and Fig.~\ref{fig:fig1}(c)  for the corresponding dynamical structure factor.
Although we focus on charge transport, energy transport is also superdiffusive following the same reasoning.

This argument can be easily generalized to higher dimensions. For the nodal interaction model in higher dimensions, the Hamiltonian is given by $H = \sum_{\bm{k}} \epsilon(\bm{k}) \hat c^\dagger_{\bm{k}} \hat c_{\bm{k}} + W\sum_{\langle i,j\rangle} \hat d^\dagger_i \hat d_i \hat d^\dagger_j \hat d_j$. An example can be found in the supplemental material~\cite{suppmat}. The corresponding Boltzmann equation is analogous to the one-dimensional case, with the one-dimensional variables $v$,$\partial_x$,$x$,and $k$ replaced by their higher-dimensional counterparts: the velocity vector $\bm v = \nabla_{\bm{k}} \epsilon(k)$, the gradient $ \nabla_{\bm r}$, the position vector $\bm r$, and the momentum vector $\bm k$. In $d$ dimensions, assuming the velocity at the nodes is non-zero, a $m$-dimensional nodal surface of order $n$ yields a dynamical exponent $z=(2n+d-m)/2n$, which only holds true for $z<2$. For example, nodal lines of order $n$ in two-dimensional systems results in $z=(2n+1)/2n$, and nodal points of order $n$ in two-dimensional systems results in $z=(n+1)/n$~\cite{Wang23}.

\textit{Numerics.---}
To check our results numerically for one-dimensional fermionic chains, we use tensor network (matrix-product operator) techniques~\cite{SCHOLLWOCK201196}. We first consider a system with boundary driving, and analyze the scaling of the current in the non-equilibrium steady-state (NESS), a method frequently employed to study transport properties~\cite{boundarydrivenRMP,znidaric11,Prosen_2009,znidaric16}. In this setup, the first and last sites of the system are coupled to external baths, modeled phenomenologically by four Lindblad operators:
\begin{align}
\begin{aligned}
	L_1=\sqrt{\Gamma(1+\mu)} \hat c^\dagger_1,\quad & L_2=\sqrt{\Gamma(1-\mu)} \hat c_1 \\
	L_3=\sqrt{\Gamma(1-\mu)} \hat c^\dagger_L,\quad & L_4=\sqrt{\Gamma(1+\mu)} \hat c_L,
\end{aligned}
\end{align}
where $\rho$ is density matrix and $\mathcal{L}^{(\text{bath})}(\rho)=\sum_{k=1}^4 2 L_k \rho L_k^{\dagger} -\rho L_k^{ \dagger} L_k-L_k^{ \dagger} L_k \rho$ . The evolution of the density matrix is then governed by the Lindblad master equation $\frac{\mathrm{d} \rho}{\mathrm{d} t}=\mathrm{i}[\rho, \hat H]+\mathcal{L}^{(\text {bath})}(\rho)$.
The NESS current scales with system size as $j\sim L^{-(z-1)}$, see {\it e.g.}~\cite{znidaric16,boundarydrivenRMP}.

We employ the time-evolving block decimation (TEBD) algorithm~\cite{Prosen_2009,PhysRevLett.91.147902,PhysRevLett.93.207204,PhysRevLett.93.207205} to obtain the NESS of the Lindblad master equation~\cite{suppmat}. The interaction strength in the Hamiltonian (Eq.~\ref{eq:Ham}) is set to a uniform value, $W_i=4$. For simplicity in simulations, we choose $\epsilon(k) = -\cos k$, $\hat{d}_i = (\hat{c}_i + ia\hat{c}_{i+1})/\sqrt{1+|a|^2}$, and correspondingly $\tilde{\phi}_k=1+ae^{i(k+\pi/2)}$.  When $a=1$, $\tilde{\phi}_k$ exhibits a node at $k_0 = \pi/2$ with order $n=1$. We present numerical results for this model (Fig.~\ref{fig:fig2}(a)). In the large $L$ limit, the scaling of the NESS current with system size follows $j\sim L^{-1/2}$ at large $L$ limit, indicating superdiffusive transport with a dynamical exponent $z=3/2$, in agreement with the predictions of the Boltzmann equation. In contrast, when $a=0.5$, $\tilde{\phi}_k\ne 0$ for all $k$, and the scaling of the NESS current satisfies $j\sim L^{-1}$ in the large $L$ limit, corresponding to diffusive transport. In our simulation, the coupling strength $\Gamma$ is set to $\Gamma=1$ and the driving parameter $\mu$ is set to be small $\mu=0.02$ to ensure we are we are in a linear response regime where all observables relevant for magnetization transport are proportional to $\mu$.

To complement these results, we also use the density matrix truncation (DMT) method~\cite{DMT} to study transport after a quantum quench from a domain wall initial state. The initial state is a domain wall state given by:
\begin{align}
    \rho(t=0) \sim (1+\nu \sigma^z)^{\otimes \frac{L}{2}}\otimes (1-\nu \sigma^z)^{\otimes \frac{L}{2}},
\end{align}
where $\hat \sigma^z = 2 \hat n -1 $ is a Pauli matrix in the spin language, using a standard Jordan-Wigner transformation. We investigate the cumulative charge flow (charge transfer) from left half to right half $\Delta s(t)\equiv \int_0^t j(x=\frac{L}{2},t') \mathrm{d}t'$. In the long-time limit, the asymptotic scaling of the cumulative charge $\Delta s$ is governed by the dynamical exponent, $\Delta s(t) \sim t^{1/z}$, see {\it e.g.}~\cite{prosen17}.
To illustrate the generality of our results and to achieve better convergence of the dynamical exponent, we choose the interaction strengths $W_i$ to be random, sampled uniformly from $[-4,4]$. Again, we use small $\nu=0.01$ to ensure we are in a linear response regime. The numerical results are presented in Fig.~\ref{fig:fig2}(b) and Fig.~\ref{fig:fig2}(c). For the nodal interaction case ($a=1$), we find $\Delta s \propto t^{2/3}$, corresponding to a dynamical exponent $z = 3/2$. Conversely, in the non-nodal interaction case ($a=0.5$), $\Delta s \propto t^{1/2}$, indicating a dynamical exponent $z = 2$.

\textit{Discussion.---}
In this work, we constructed interacting, chaotic superdiffusive models using two key elements: (1) a free-fermion Hamiltonian that possesses ballistic eigenmodes, (2) nodal interactions, where each local term commutes with the particle number operator $\hat n_{k_0}$ with a specific momentum $k_0$. These two elements lead to a asymptotically divergent lifetime for quasi-particles near momentum $k_0$. Quasi-particles with momenta near $k_0$ are responsible for the superdiffusive transport. The dynamical exponent $z$ is determined by the order $n$ of the node, and is given by $z=(2n+1)/2n$ in one dimension. 

Our work provides a counterexample to the conventional wisdom that chaotic lattice models exhibit diffusive behavior, offering a clear analytic explanation for the origin of superdiffusion. Although superdiffusion in these models relies on some degree of fine-tuning, our construction remains very general, and could provide an explanation for the observed superdiffusion in PXP and related models where isolated ballistic modes were also observed~\cite{Papic23}. We established superdiffusion in a non-perturbative way, and the interaction strength does not affect the dynamical exponent. Superdiffusion can even occur for inhomogeneous (random) interaction strengths. Furthermore, our construction is not limited to one dimension -- it is also possible to construct superdiffusive models in two and three dimensions. 

It would be very interesting to extend our results to perturbed {\em interacting} integrable systems. Integrable systems generically possess stable, ballistic quasi-particle excitations. A key challenge would be to identify local operators in such modes that commute with the quasi-particle number operator at specific momenta. Using these operators as integrability-breaking perturbations would naturally lead to superdiffusive behavior. We leave this extension for future work.

\begin{acknowledgments}
{\it Acknowledgments.} Y.-P. W. thanks Chen Fang, Marko \v Znidari\v c, Enej Ilievski and Curt von Keyserlingk for useful discussion. Y.-P. W. is supported by Chinese Academy of Sciences under grant number XDB33020000, National Natural Science Foundation of China (NSFC) under grant number 12325404, 12188101 and National Key R\&D Program of China under grant number 2022YFA1403800, 2023YFA1406704. S.G. acknowledges support from NSF QuSEC-TAQS OSI 2326767. J.R. acknowledges support by the Leverhulme Trust Research Leadership Award RL-2019-015. R.V. acknowledges partial support from the US Department of Energy, Office of Science, Basic Energy Sciences, under award No. DE-SC0023999.
The numerical simulations based on tensor networks use the ITensor package~\cite{ITensor}.
\end{acknowledgments}

\bibliography{ref}
\newpage


\section*{Supplementary material}

# Supplemental material for
# "Superdiffusive transport in chaotic systems with nodal interactions"

Yu-Peng Wang,[1,2,*] Jie Ren,[1,3,*] Sarang Gopalakrishnan,[4] and Romain Vasseur[5,6]

[1]*Beijing National Laboratory for Condensed Matter Physics and Institute of Physics,
Chinese Academy of Sciences, Beijing 100190, China*
[2]*University of Chinese Academy of Sciences, Beijing 100049, China*
[3]*School of Physics and Astronomy, University of Leeds, Leeds LS2 9JT, United Kingdom*
[4]*Department of Electrical and Computer Engineering,
Princeton University, Princeton, NJ 08544, USA*
[5]*Department of Theoretical Physics, University of Geneva,
24 quai Ernest-Ansermet, 1211 Genève, Switzerland*
[6]*Department of Physics, University of Massachusetts, Amherst, MA 01003, USA*

(Dated: February 7, 2025)

## CONTENTS

## I. LOWER BOUND OF DIFFUSION CONSTANT

The diffusion constant at infinite temperature can be expressed through the current-current correlation function as:

$$D = 4 \lim_{t \to \infty} \lim_{L \to \infty} \frac{1}{L} \int_0^t d\tau \langle \hat{J}(\tau) \hat{J}(0) \rangle, \tag{1}$$

where $\langle \cdots \rangle = V^{-1} \text{Tr}(\cdots)$, $V$ is dimension of Hilbert space, and $\hat{J}(t)$ represents the total current, defined as the sum of local currents across all lattice sites $\hat{J}(t) = \sum_r \hat{j}_r(t)$.

Hydrodynamically projecting the current onto the occupation numbers $\hat{n}_k$ with $k \in \mathcal{K}(t) = (k_0 - \delta k(t), k_0 + \delta k(t))$, we obtain the slow component $\hat{J}_s$ of current, which remains conserved up to time $t$. The remaining fast component $\hat{J}_f = \hat{J} - \hat{J}_s$ is orthogonal to the slow part. The current-current correlation can then be expressed as:

$$\langle \hat{J}(\tau) \hat{J}(0) \rangle = \langle \hat{J}_s(\tau) \hat{J}_s(0) \rangle + \langle \hat{J}_f(\tau) \hat{J}_f(0) \rangle + \langle \hat{J}_s(\tau) \hat{J}_f(0) \rangle + \langle \hat{J}_f(\tau) \hat{J}_s(0) \rangle. \tag{2}$$

Since $J_s$ is conserved and orthogonal to $J_f$, we have $\langle \hat{J}_s(\tau) \hat{J}_f(0) \rangle = \langle \hat{J}_s(0) \hat{J}_f(0) \rangle = 0$ and $\langle \hat{J}_f(\tau) \hat{J}_s(0) \rangle = \langle \hat{J}_f(\tau) \hat{J}_s(\tau) \rangle = 0$. Additionally, the contribution from the fast part $\hat{J}_f$ is positive:

$$\begin{aligned}
&\lim_{t \to \infty} \lim_{L \to \infty} \int_0^t \langle \hat{J}_f(\tau) \hat{J}_f(0) \rangle d\tau \\
&= \lim_{\Omega \to 0} \lim_{t \to \infty} \lim_{L \to \infty} \int_0^t \sum_{m,n} \left| \langle m | \hat{J}_f(0) | n \rangle \right|^2 e^{i(E_m - E_n)\tau} e^{-\tau^2 \Omega^2} d\tau \\
&= \lim_{\Omega \to 0} \lim_{L \to \infty} \sum_{m,n} e^{-(E_m - E_n)^2 / 4\Omega^2} \left| \langle m | \hat{J}_f(0) | n \rangle \right|^2 > 0,
\end{aligned} \tag{3}$$

where $|m\rangle$ ($|n\rangle$) denotes the eigenstate of Hamiltonian $\hat{H}$ with the corresponding eigenvalue $E_m$ ($E_n$), and $\Omega$ serves as a filter parameter to prevent singularities when altering the order of limits.

Thus, the diffusion constant has the following lower bound:

$$D > 4 \lim_{t \to \infty} \lim_{L \to \infty} \frac{1}{L} \int_0^t d\tau \langle \hat{J}_s^2 \rangle \equiv D_s. \tag{4}$$

## II. LONG-LIVED MODES $\hat{n}_k$

Here, we aim to analyze the size of the subset $\mathcal{K} \equiv (k_0 - \delta k(t), k_0 + \delta k(t))$, where the occupation number $n_k$ remains approximately conserved up to time $t$. For this purpose, we utilize the inequality introduced in Ref. 1:

$$|\langle (\hat{O}(t) - \hat{O}(0))\hat{O}(0) \rangle| \leq \langle [\hat{H}, \hat{O}]^2 \rangle t^2/2, \tag{5}$$

where $\langle \hat{A} \rangle = 2^{-L} \text{Tr}(\hat{O})$ is expectation value of operator $\hat{A}$ in a size $L$ system at infinite temperature. $\hat{O}$ is an arbitrary operator. Here, we apply this inequality to $\hat{O} = \hat{n}_k$, considering the Hamiltonian with nodal interaction: $\hat{H} = \hat{H}_0 + \hat{V} = \sum_k \epsilon_k \hat{c}_k^\dagger \hat{c}_k + W \sum_i \hat{d}_i^\dagger \hat{d}_i \hat{d}_{i+1}^\dagger \hat{d}_{i+1}$. When $\langle (\hat{n}(t) - \hat{n}(0))\hat{n}(0) \rangle < \epsilon$ with a small $\epsilon$ for all times $\tau \in (0, t)$, the $\hat{n}_k$ can be considered approximately conserved up to time $t$ at infinite temperature.

Next, we need to analyze the expectation value of commutator $\langle [\hat{H}, \hat{O}]^2 \rangle$. The free fermion Hamiltonian $\hat{H}_0$ commutes with $\hat{n}_k$, so we only need to consider the commutator with the interaction term $\hat{V}$. The interaction can be Fourier-transformed into momentum space as $\hat{V} = \frac{1}{L} \sum_{k_1, k_2, q} U_{k_1, k_2, k_1+q, k_2-q} \hat{c}_{k_1}^\dagger \hat{c}_{k_1+q} \hat{c}_{k_2}^\dagger \hat{c}_{k_2-q}$, where $U_{k_1, k_2, k_1+q, k_2-q} = d_{k_1}^* d_{k_2}^* d_{k_1+q} d_{k_2-q} \cos q$ and $L$ is system size. The Fourier-transformed fermion operators are defined as $\hat{c}_k = \frac{1}{\sqrt{L}} \sum_x e^{ikx} \hat{c}_x$.

The commutator $\left[\hat{H}, \hat{n}_k\right]$ is given by:

$$\begin{aligned}
\left[\hat{H}, \hat{n}_k\right] =& \frac{1}{L} \sum_{k_1, k_2, q} d_{k_1}^* d_{k_2}^* d_{k_1+q} d_{k_2-q} \cos q\, \hat{c}_{k_1}^\dagger \hat{c}_{k_1+q} \hat{c}_{k_2}^\dagger \hat{c}_{k_2-q} (\delta_{k,k_2-q} + \delta_{k,k_1+q} - \delta_{k,k_2} - \delta_{k,k_1}) \\
=& \frac{1}{L} d_k \sum_{k_1, q} d_{k+q}^* d_{k_1-q}^* d_{k_1} \cos q (\hat{c}_{k+q}^\dagger \hat{c}_k \hat{c}_{k_1-q}^\dagger \hat{c}_{k_1} + \hat{c}_{k_1-q}^\dagger \hat{c}_{k_1} \hat{c}_{k+q}^\dagger \hat{c}_k) \\
& - \frac{1}{L} d_k^* \sum_{k_1, q} d_{k_1}^* d_{k_1+q} d_{k-q} \cos q (\hat{c}_{k_1}^\dagger \hat{c}_{k_1+q} \hat{c}_k^\dagger \hat{c}_{k-q} + \hat{c}_k^\dagger \hat{c}_{k-q} \hat{c}_{k_1}^\dagger \hat{c}_{k_1+q}),
\end{aligned} \tag{6}$$

where each term is proportional to $d_k$ or $d_k^*$. Thus, the square of the commutator can be expressed in the form: $[\hat{H}, \hat{n}_k]^2 = |d_k|^2 \frac{1}{L^2} \sum_{k_1 \ldots k_8} U_{k_1 \ldots k_8} \hat{c}_{k_1}^\dagger \hat{c}_{k_2} \hat{c}_{k_3}^\dagger \hat{c}_{k_4} \hat{c}_{k_5}^\dagger \hat{c}_{k_6} \hat{c}_{k_7}^\dagger \hat{c}_{k_8}$. The coefficients $U_{k_1 \ldots k_8}$ are nonzero only if: (1) momentum conservation is satisfied, i.e. $k_1 + k_2 = k_3 + k_4$ and $k_5 + k_6 = k_7 + k_8$. (2) one of $\{k_1, k_2, k_3, k_4\}$ and one of $\{k_5, k_6, k_7, k_8\}$ are equal to $k$. These two conditions reduce the summation to only 4 independent indices. Additionally, the nonzero trace requires that the momenta of creation and annihilation operators match in pairs, leaving only $L^2$ terms contribute to $\langle [\hat{H}, \hat{n}_k]^2 \rangle$, which precisely cancel the $L^{-2}$ factor outside the sum, yielding:

$$\langle [\hat{H}, \hat{n}_k]^2 \rangle = c^2 |d_k|^2, \tag{7}$$

with an $O(1)$ constant $c$.

Thus, when $c|d_k|^2 t^2/2 < \epsilon$, the occupation number $n_k$ remains approximately conserved up to time $t$. For an order $n$ node, i.e. $d_{k_0+q} \sim q^n$, the $\hat{n}_k$ conserved subset $\mathcal{K} \equiv (k_0 - \delta k(t), k_0 + \delta k(t))$ satisfies the scaling relation $\delta k \sim t^{-1/n}$.

## III. COLLISION FUNCTION FOR GENERAL $m$-BODY SCATTERING

In this part, we derive the $m$-body scattering collision term in the quantum Boltzmann equation for the model with nodal interactions, and show that it can be expressed in the general form introduced in Eq. (4) of the main text:

$$f_k^{\text{col}}[\delta n] = -\gamma(k)|d_k|^2 \delta n(x, k, t) + \gamma(k)|d_k|^2 \int dk_1\, \beta(k, k_1) \delta n(x, k_1, t), \tag{8}$$

We can write the Hamiltonian in momentum space as:

$$H = \int \frac{dk}{2\pi}\epsilon(k)\hat{c}_k^\dagger \hat{c}_k + \int \frac{dk^4}{(2\pi)^3}\delta_{2\pi}(\underline{k})U_{k_1,k_2,k_1',k_2'}\hat{c}_{k_1}^\dagger \hat{c}_{k_1'}\hat{c}_{k_2}^\dagger \hat{c}_{k_2'}, \tag{9}$$

where $U_{k_1,k_2,k_1',k_2'} = W d_{k_1}^* d_{k_2}^* d_{k_1'} d_{k_2'} \cos(k_1 - k_1')$.

The Boltzmann equation can be written as $\partial_t n_k(x,t) = -v(k)\partial_x n_k(x,t) + f_{\text{col}}(n_k(x,t))$, where the collision function contains collision of different number of particles $f_k^{\text{col}}[\delta n] = \sum_m f_k^{\text{col},(m)}[\delta n]$. The m-particle collision integral reads

$$f_{k_1}^{\text{col},(m)}[\delta n] = \frac{4\pi}{\hbar}\int \frac{dk_2\ldots dk_m dk'\ldots dk_m}{(2\pi)^{2m-1}} W_{k_1,k_2,\ldots,k_m;k_1',k_2',\ldots,k_m'} \tag{10}$$
$$\times [(1-n_{k_1})\ldots(1-n_{k_m})n_{k_1'}\ldots n_{k_m'} - (1-n_{k_1'})\ldots(1-n_{k_m'})n_{k_1}\ldots n_{k_m}],$$

where $W_{k_1,k_2,\ldots,k_m;k_1',k_2',\ldots,k_m'}$ is the scattering probability.

At thermal equilibrium with infinite temperature, we have $f_{\text{col}}[n_{\text{eq}}(x,k,t)] = 0$ and $n_{\text{eq}}$ is independent of $x$ and $k$. When $n_k(x,t)$ deviates slightly from equilibrium, such that $n_k(x,t) = n_{\text{eq}} + \delta n_k(x,t)$, the linearized collision function can be obtained by neglecting higher-order corrections of $O(\delta n^2)$

$$f_{k_1}^{\text{col},(m)}[\delta n] = -\frac{4\pi}{\hbar}\sum_l \int \prod_{i=1}^m \frac{dk_i'}{2\pi} \prod_{j=2}^m \frac{dk_j}{2\pi} W_{k_1,k_2,\ldots,k_m;k_1',k_2',\ldots,k_n'}(1-n_{\text{eq}})^{m-1} n_{\text{eq}}^{m-1} \delta n_{k_l}$$
$$+ \frac{4\pi}{\hbar}\sum_l \int \prod_{i=1}^m \frac{dk_i'}{2\pi}\prod_{j=2}^m \frac{dk_j}{2\pi} W_{k_1,k_2,\ldots,k_m;k_1',k_2',\ldots,k_m'}(1-n_{\text{eq}})^{m-1} n_{\text{eq}}^{m-1}\delta n_{k_l'}. \tag{11}$$

Since the scattering probability satisfies $W_{k_1,k_2,\ldots,k_m;k_1',k_2',\ldots,k_m'} = W_{k_1',k_2',\ldots,k_m';k_1,k_2,\ldots,k_m}$, the positive and negative terms cancel out except for the case $l=1$.

Using the generalized Fermi's golden rule, the scattering probability can be derived from inserting T-matrix, $T = V + VG_0V + VG_0VG_0V + \cdots$. The $m$-particle scattering probability is

$$W_{k_1,k_2,\ldots,k_m;k_1',k_2',\ldots,k_m'} = |\langle k_1,k_2,\ldots,k_m| \overbrace{VG_0V\cdots G_0V}^{\text{the number of V is } m-1} |k_1',k_2',\ldots,k_m'\rangle|^2 \delta(\underline{\epsilon}), \tag{12}$$

where $\underline{\epsilon} = \sum_i (\epsilon_{k_i} - \epsilon_{k_i'})$ and

$$\langle k_1,k_2,\ldots,k_m| \overbrace{VG_0V\cdots G_0V}^{\text{the number of V is } m-1} |k_1',k_2',\ldots,k_m'\rangle = \sum_{(a_1\ldots a_m)\in(k_1\ldots k_m)} \sum_{(a_1'\ldots a_m')\in(k_1'\ldots k_m')} \text{sgn}(a_1\ldots a_m)\text{sgn}(a_1'\ldots a_m')$$
$$\times \delta(\underline{a})\prod_{i=1}^{m-1} U_{a_i,a_{i+1},a_i',a_{i+1}'} \prod_{i=2}^{m-1} \frac{1}{(\epsilon_{a_i}+\epsilon_{a_{i+1}}-\epsilon_{a_i'}-\epsilon_{a_{i+1}'}+i\eta)}. \tag{13}$$

Since $U_{a_i,a_{i+1},a_i',a_{i+1}'} \propto d_{a_i}^* d_{a_{i+1}}^* d_{a_i'} d_{a_{i+1}'}$, $W_{k_1,k_2,\ldots,k_m;k_1',k_2',\ldots,k_m'}$ can be factorized as $W_{k_1,k_2,\ldots,k_m;k_1',k_2',\ldots,k_m'} = A(k_1,k_2,\ldots,k_m;k_1',k_2',\ldots,k_m')\Pi_{i=1}^m |d_{k_i}|^2 |d_{k_i'}|^2$, where $A(k_1,k_2,\ldots,k_m;k_1',k_2',\ldots,k_m')$ is nonnegative for all parameters. Then we can write the linearized collision function as

$$f_{k_1}^{\text{col},(m)}[\delta n] = -\frac{4\pi}{\hbar}|d_{k_1}|^2(1-n_{\text{eq}})^{m-1}n_{\text{eq}}^{m-1}\delta n_{k_1}\int \prod_{a=1}^m (\frac{dk_a'}{2\pi}|d_{k_a'}|^2)\prod_{j=2}^m (\frac{dk_j}{2\pi}|d_{k_j}|^2) A_{k_1,k_2,\ldots,k_m;k_1',k_2',\ldots,k_m'}$$
$$+\frac{4\pi}{\hbar}|d_{k_1}|^2(1-n_{\text{eq}})^{m-1}n_{\text{eq}}^{m-1}\int \prod_{a=1}^m (\frac{dk_a'}{2\pi}|d_{k_a'}|^2)\prod_{j=2}^m (\frac{dk_j}{2\pi}|d_{k_j}|^2) A_{k_1,k_2,\ldots,k_m;k_1',k_2',\ldots,k_m'}\delta n_{k_1'}. \tag{14}$$

To express the collision function in the general form mentioned in Eq. 8, we define the following:

$$\gamma_m(k_1) = \frac{4\pi}{\hbar}(1-n_{\text{eq}})^{m-1}n_{\text{eq}}^{m-1}\int \prod_{a=1}^m (\frac{dk_a'}{2\pi}|d_{k_a'}|^2)\prod_{j=2}^m (\frac{dk_j}{2\pi}|d_{k_j}|^2) A_{k_1,k_2,\ldots,k_m;k_1',k_2',\ldots,k_m'}$$
$$\beta_m(k_1,k_1') = \frac{2|d_{k_1'}|^2}{\gamma_m \hbar}(1-n_{\text{eq}})^{m-1}n_{\text{eq}}^{m-1}\int \prod_{a=2}^m (\frac{dk_a'}{2\pi}|d_{k_a'}|^2)\prod_{j=2}^m (\frac{dk_j}{2\pi}|d_{k_j}|^2) A_{k_1,k_2,\ldots,k_m;k_1',k_2',\ldots,k_m'}. \tag{15}$$

With these definitions, the linearized collision function can be expressed as:

$$f_k^{\text{col,(m)}}[\delta n] = -\gamma_m(k)|d_k|^2 \delta n_k(x,t) + \gamma_m(k)|d_k|^2 \int dk_1 \beta_m(k_1,k) \delta n_{k_1}(x,t), \tag{16}$$

The total collision function can then be written in the general form as in Eq. 8:

$$f_k^{\text{col}}[\delta n] = \sum_m f_k^{\text{col,(m)}}[\delta n] = -\gamma(k)|d_k|^2 \delta n_k(x,t) + \gamma(k)|d_k|^2 \int dk_1 \beta(k_1,k) \delta n_{k_1}(x,t), \tag{17}$$

where $\gamma(k) = \sum_{m=2}^{+\infty} \gamma_m(k)$ and $\beta(k_1,k) = \frac{\sum_{m=2}^{+\infty} \gamma_m(k)\beta_m(k_1,k)}{\gamma(k)}$. Since all $\gamma_m$ are nonnegative and, in general, not all $\gamma_m$ are 0 at the same $k$, we have $\gamma(k) > 0$ for all $k$.

## IV. CURRENT OF NON-EQUILIBRIUM STEADY STATE

We follow the methodology outlined in Ref. 2 to study the non-equilibrium steady state (NESS) current of our model in the boundary-driven setup.

This set-up is described by the Lindblad master equation $\frac{d\rho}{dt} = i[\rho, \hat{H}] + \mathcal{L}^{(\text{bath})}(\rho)$, where $\mathcal{L}^{(\text{bath})}(\rho) = \sum_{k=1}^{4} 2L_k \rho L_k^\dagger - \rho L_k^\dagger L_k - L_k^\dagger L_k \rho$ represents the coupling between the system boundaries and external baths, and the Lindbald operators are given by:

$$\begin{aligned} L_1 &= \sqrt{\Gamma(1+\mu)}\hat{c}_1^\dagger, & L_2 &= \sqrt{\Gamma(1-\mu)}\hat{c}_1 \\ L_3 &= \sqrt{\Gamma(1-\mu)}\hat{c}_L^\dagger, & L_4 &= \sqrt{\Gamma(1+\mu)}\hat{c}_L. \end{aligned} \tag{18}$$

This coupling to external baths fixes a particle number imbalance between the two ends of the system, which drives a current in the NESS. The NESS current scales with system size as $j \sim L^{-(z-1)}$.

To employ tensor network (matrix-product operator) techniques for solving the NESS, we first perform a standard Jordan-Wigner transformation to map the system into the spin representation. The density matrix $\rho$ is expressed as a linear combination of all possible products of local Pauli operators, which form a complete basis for the $4^n$-dimensional Hilbert space of operators:

$$|\rho\rangle = \sum_{\underline{s}} c_{\underline{s}} |\sigma^{\underline{s}}\rangle, \tag{19}$$

where we use the shorthand notation $\sigma^{\underline{s}} \equiv \sigma_1^{s_1} \otimes \cdots \otimes \sigma_L^{s_L}$, $\underline{s} \equiv \{s_1, \cdots, s_n\}$, and $s_i \in \{0,1,2,3\}$. Here, $\sigma^0 = \mathbb{I}$, $\sigma^1 = \sigma^x$, $\sigma^2 = \sigma^y$, $\sigma^3 = \sigma^z$ represent the identity operator and the Pauli matrices, respectively. The subscript of each Pauli matrix indicates its spatial position in the lattice. We represent the coefficients in a matrix product form as $c_{\underline{s}} = \text{Tr}(\mathbf{A}_1^{s_1} \cdots \mathbf{A}_n^{s_n})$, where each $\mathbf{A}_i^{s_i}$ is an $N \times N$ matrix. $N$ is bond dimension, which is set to 400 in our simulations.

We initialize the system in a maximally mixed state and follow the standard time-evolving block decimation (TEBD) procedure to perform the time evolution. After a sufficiently long time, the state $\hat{\rho}(t)$ converges to the NESS, denoted as $\hat{\rho}_s$. The NESS current is calculated as $j = \text{Tr}(\hat{j}\hat{\rho}_s)$, where the current operator $\hat{j}$ is derived from the continuity equation $\frac{d\hat{n}_m}{dt} = \hat{j}_{m-1} - \hat{j}_m$.

---



\end{document}